\documentclass{jps-cp}

\title{Large Area Photo-Detection System using 3-inch PMTs for the Hyper-Kamiokande Outer Detector}

\author{Stephane \textsc{Zsoldos}$^{1}$}

\inst{$^{1}$Queen Mary University of London, 327 Mile End Road, E1 4NS, London, United Kingdom}

\email{s.zsoldos@qmul.ac.uk}

\recdate{March 14, 2019}

\abst{
Hyper-Kamiokande, scheduled to begin construction as soon as 2020, is a next
generation underground water Cherenkov detector, based on the highly successful
Super-Kamiokande experiment. It will serve as a far detector, 295~km away, of a
long baseline neutrino experiment for the upgraded J-PARC beam in Japan. It will
also be a detector capable of observing --- far beyond the sensitivity of the
Super-Kamiokande detector --- proton decay, atmospheric neutrinos, and neutrinos
from astronomical sources.

An Outer Detector (OD) consisting of PMTs mounted behind the inner detector PMTs
and facing outwards to view the outer shell of the cylindrical tank,  would
provide topological information to identify interactions originating from
particles outside the inner detector. Any optimization would lead to a
significant improvement for the physics goals of the experiment, which are the
measurement of the CP leptonic phase and the determination of the neutrino mass
hierarchy.

An original setup using small 3” PMTs is being designed for the Hyper-Kamiokande
OD. They would give better redundancy, spatial, and angular resolution, as they
would be twice or three times more photosensors that the original 8” design
proposal of the experiment, and for a reduce cost. Several 3” PMTs candidates
considered for the Hyper-Kamiokande OD have been characterized at Queen Mary
University London. They all show a very low dark counts and good collection
efficiency, which makes them excellent choice to be used in the experiment.
}

\kword{3" photosensors, neutrino oscillation experiment, Hyper-Kamiokande}

\begin{document}
\maketitle

\section{Introduction}

Hyper-Kamiokande\cite{Abe:2018uyc} is the succesor of the
Super-Kamiokande\cite{Fukuda:2002uc} experiment which was awarded with the Nobel
Prize in 2015 for the joint discovery with the SNO experiment of atmospheric
neutrino oscillations. Compared to Super-Kamiokande, this new experiment will
consists of two cylindrical water tanks that are 60\,m in height and 74\,m in
diameter, for a total volume up to 0.5\,Mm$^3$ per tank.

Hyper-Kamiokande will be a multipurpose neutrino detector with a rich physics
program that aims to address some of the most significant questions facing
particle physicists today. Oscillation studies from accelerator, atmospheric and
solar neutrinos will refine the neutrino mixing angles and mass squared
difference parameters and will aim to make the first observation of asymmetries
in neutrino and antineutrino oscillations arising from a CP-violating phase,
shedding light on one of the most promising explanations for the
matter-antimatter asymmetry in the Universe. The search for nucleon decays will
probe one of the key tenets of Grand Unified Theories. In the case of a nearby
supernova, Hyper-Kamiokande will observe an unprecedented number of neutrino
events, providing much needed experimental results to researchers seeking to
understand the mechanism of the explosion. Finally, the detection of
astrophysical neutrinos from sources such as dark matter annihilation, gamma ray
burst jets, and pulsar winds could further our understanding of some of the most
spectacular, and least understood, phenomena in the Universe.

\begin{figure}[!htb]
  \begin{minipage}[c]{0.49\textwidth}
    \centering
    \includegraphics[width=1.\textwidth]{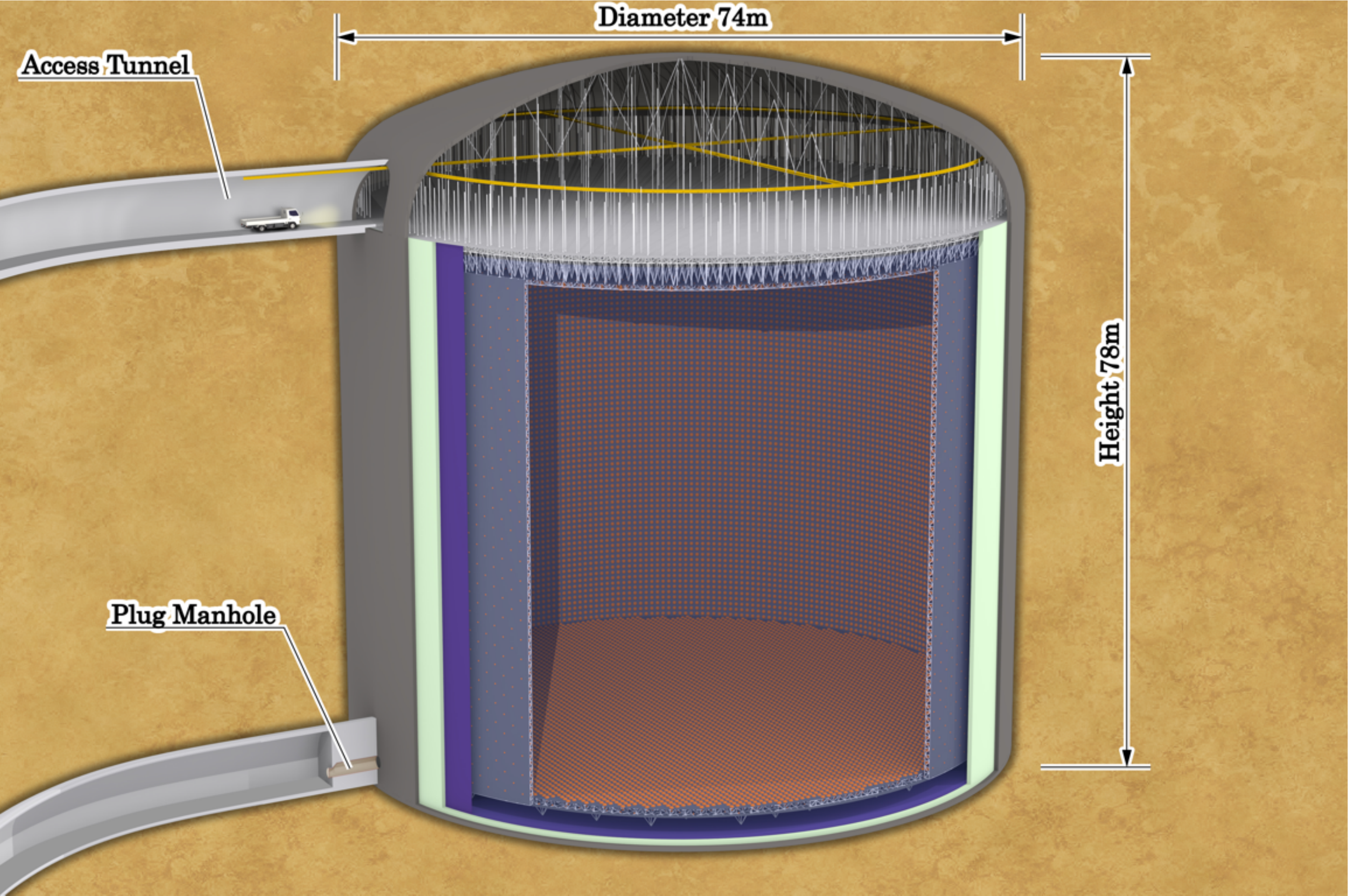}
  \end{minipage}\hfill
  \begin{minipage}[c]{0.49\textwidth}
    \centering
    \includegraphics[width=1.\textwidth]{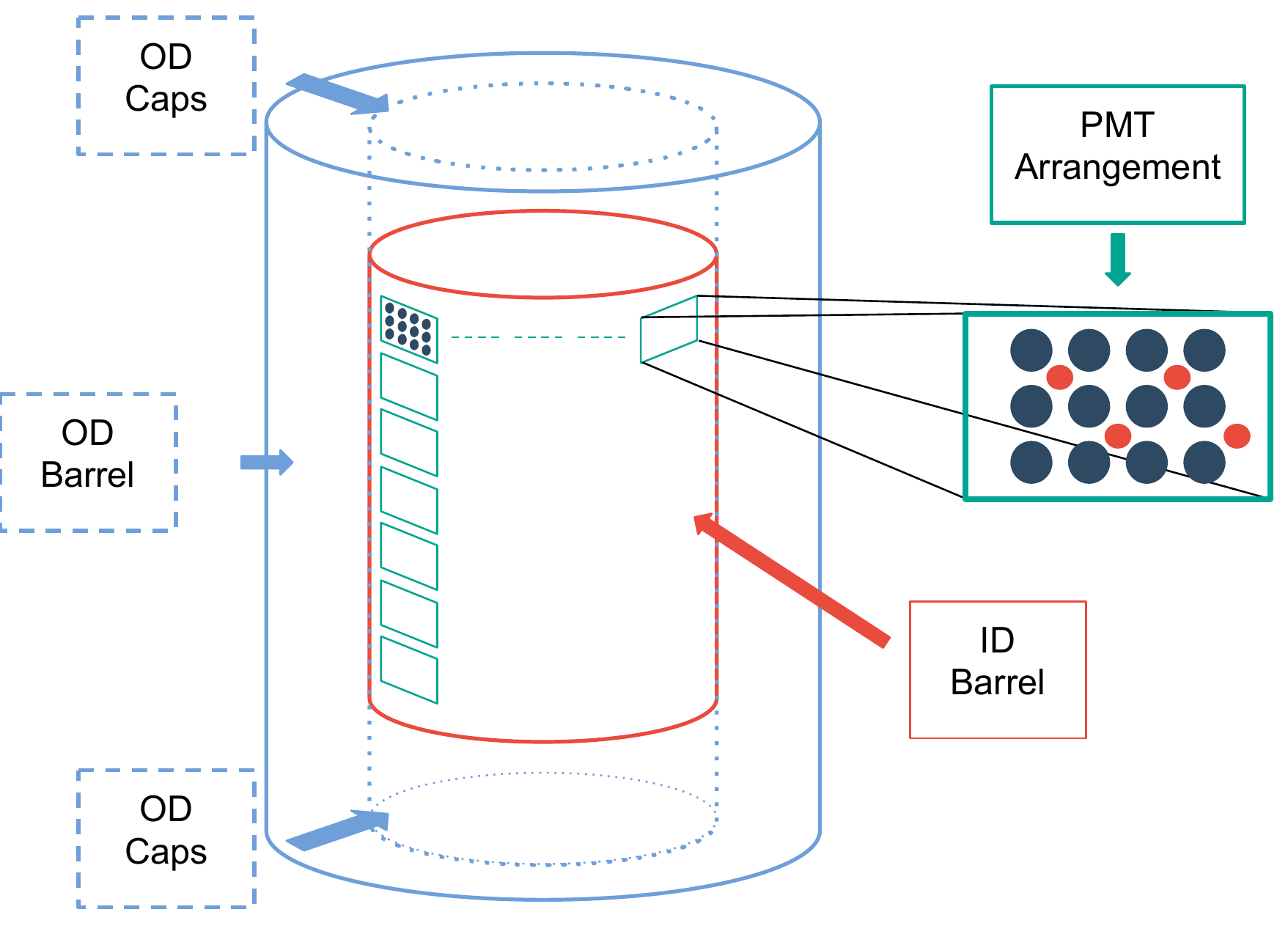}
  \end{minipage}
  \caption{
   Left\,: Schematic view for the configuration of single cylindrical tank
   instrumented with high density (40\% photocoverage) PMTs. Right\,: A sketch
   of the Hyper-Kamiokande detector design (not to scale). The structure holding
   the ID photosensors is represented in red, with the limits of the tank in
   blue. The area outside the ID detector is the OD volume, where we distinguish
   the barrel region from the top/bottom endcaps. The photosensors are arranged
   with respect to the green rectangle on the right side of the figure. The dark
   blue photosensors are the ID ones, and the red ones are the OD ones, facing
   outwards. The OD photosensors shown correspond to a total of 13.3k 3" PMTs.
   }
  \label{HK1Tank}
\end{figure}

Hyper-Kamiokande employs a ring-imaging water Cherenkov detector technique and
as a consequence, the capability of a water Cherenkov detector largely relies on
the performance of its photosensors. The detector represented in
Fig.\ref{HK1Tank} is segmented in two parts, an inner-detector (ID) surrounded
by 40000 20" photomultipliers (PMTs), and an outer-detector (OD) with 13300 3"
PMTs aimed to reject background from any source.

Neutrino interactions are characterised by a lack of incoming particles, and it
is important to veto events where there is activity in the outer part of the
detector. Low energy neutrino interactions produce signals that can be swamped
by background from low energy (1 to 10\,MeV) gammas and neutrons. These
backgrounds are partly due to natural radioactivity in the surrounding rock, and
in the photodetectors themselves, but there is also a contribution from
spallation interactions by cosmic muons. The reconstruction of events uses the
expected Cherenkov cone pattern from a charged particle, and the addition of
background photons leads to mis-reconstruction and misidentification of the
particles.

The second source of background is the hard component of the cosmic muon which
penetrates deep inside the Earth. Muons that enter the outer detector create a
large number of Cherenkov photons which can be identified by the outer
photosensors. A very efficient veto against incoming muons is essential for the
physics programme, particularly for atmospheric neutrinos and proton decay
searches.

To veto activity in the OD it has to be optically separated from the ID, with
photons detected by a separate array of photosensors.  The current design has an
outer layer thickness between 1\,m (barrel) and 2.5\,m (endcap), and a dead
region of 60\,cm between the OD and ID photosensors, determined by the size of
the ID covers. From Super-Kamiokande we know that this is sufficient to contain
most, but not all the low energy backgrounds.

\section{Outer Detector Photosensors}

Based on the experience acquired on the Super-Kamiokande experiment, several
criterias has been defined to select the photosensors candidates. The idea
behind using a large array of small PMTs compared to larger ones in this
previous experiment comes from the way the OD hits information are used. The OD
is a veto for background particles, which is based on cluster of photosensors
hits above a certain threshold, defined by the photosensors dark counts.
Therefore, the OD segment is used as a photon counter, compared to the ID where
the event topology is also used to reconstruct the neutrino energy and vertex
position. In the OD case, the amount of information contains in an event
increase linearly with the number of PMTs, as defined by Eq.\ref{eq:ODentropy}
the information entropy of our signal :

\begin{equation}
  H = - \log_2{2^{-N}} \hspace{1em} \text{with N = Number of PMTs}
  \label{eq:ODentropy}
\end{equation}

Hence, three criterias emerges for building an efficient OD for
Hyper-Kamiokande\,: we have to increase the number of photosensors, achieved by
using smaller and inexpensive 3" PMTs, with a small dark rate to lower the
trigger threshold and of course a good light collection, to set an efficient
trigger to background events.

\begin{figure}[!htb]
  \begin{minipage}[c]{0.21\textwidth}
    \centering
    \includegraphics[width=1.\textwidth]{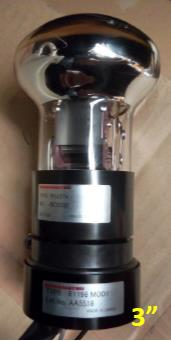}
  \end{minipage}\hfill
  \begin{minipage}[c]{0.30\textwidth}
    \centering
    \includegraphics[width=1.\textwidth]{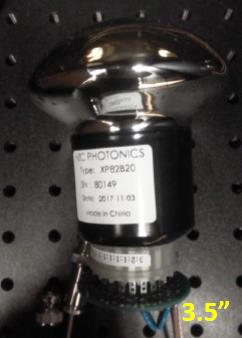}
  \end{minipage}\hfill
  \begin{minipage}[c]{0.33\textwidth}
    \centering
    \includegraphics[width=1.\textwidth]{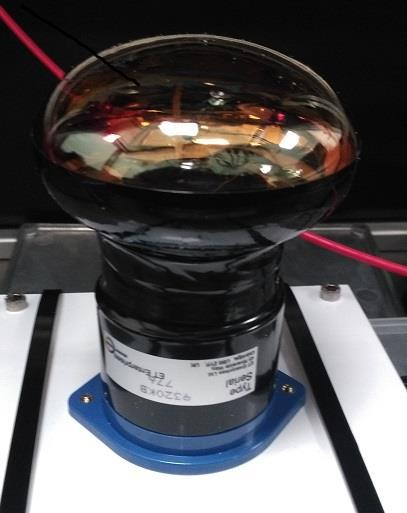}
  \end{minipage}
\caption{
  Left\;: Hamamatsu 3" R14374.
  Center\;: HZC 3.5" XP82B20.
  Right\;: ETEL 3.5" ET9320KFL.
}
\label{PMTs}
\end{figure}

In light of this, several PMTs candidates have been selected to be
characterised\,: two Hamamatsu PMTs, the 3" R14374 and the 3.5" R14689; one
Electron Tubes 3.5" PMT model ET9320KFL; and one HZC 3.5" PMT model XP82B20.
These PMTs are shown in pictures Fig.\ref{PMTs}. The 3" R14374 Hamamatsu is the
same PMT used in the Km3NeT experiment\cite{Aiello:2018nvl}, otherwise the other
3.5" candidates are in development and in need of accurate measurements from
various sources.

\begin{figure}[!htb]
  \centering
    \includegraphics[width=0.80\linewidth]{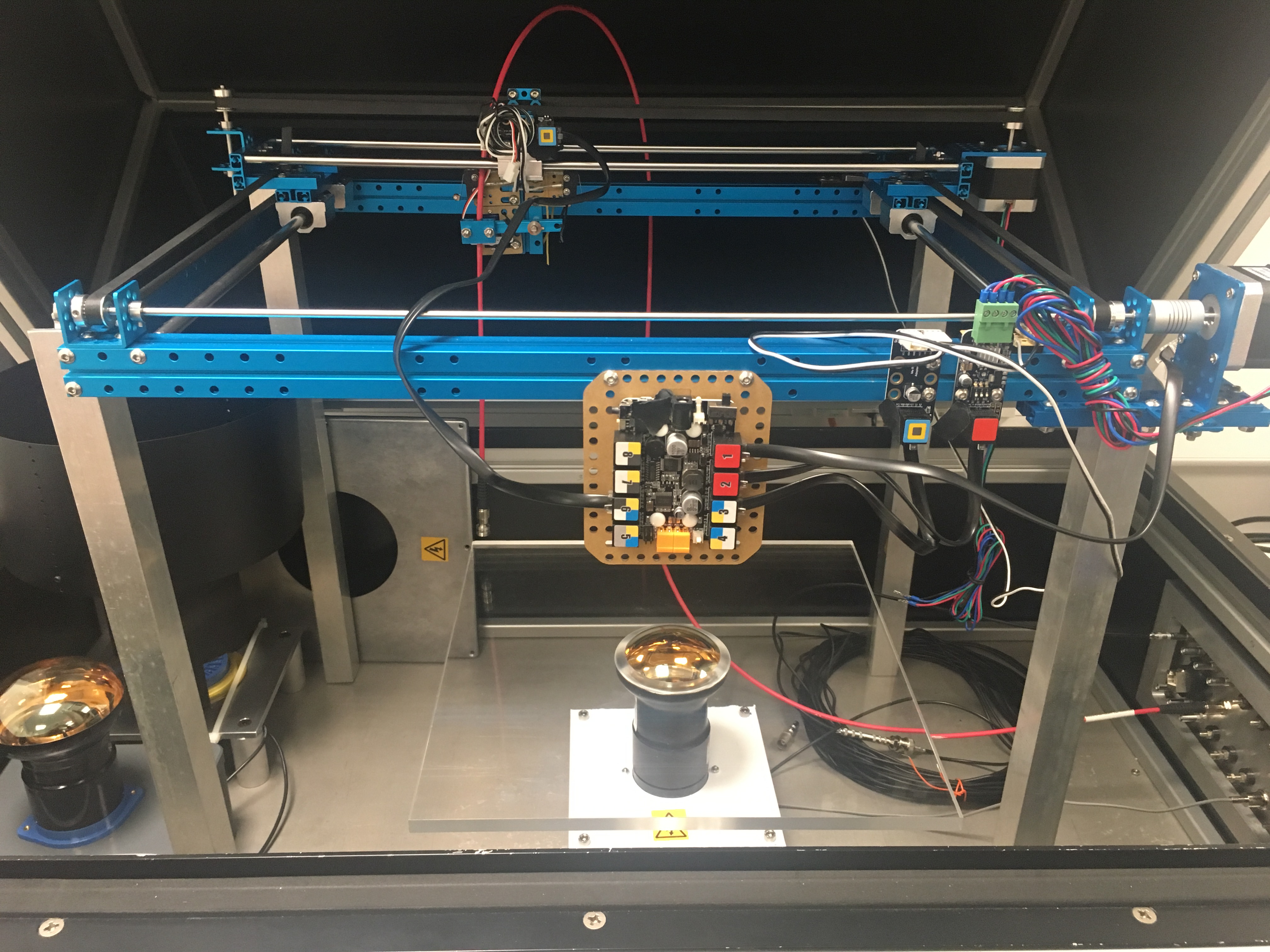}
    \caption{
    Black box for photosensors testing at Queen Mary. This picture was taken
    with a Hamamatsu 3" R14374 PMT. One can see the XY stage above the PMT
    moving along the X-Y axis an optical fibre (in yellow), guiding the light
    out from a LED driver to the blackbox.
    }
    \label{fig:setup}
\end{figure}

The results described in this proceeding have been performed at Queen Mary
University of London, using a setup shown in Fig.\ref{fig:setup}. It consists of
a black box where the 3" PMT lies inside, an automated X-Y stage controlled by
an arduino which moves an LED powered by a driver adjusting the light emitted, a
CAEN SP5601 module with a OSSV5111A high power LED, with a range from a few to
tens of thousands photons. A VME SIS6136 is in charge of the digitization of the
data, sampled with a 12-bit FADC at 250\,MHz and recorded on 1024 samples
(4096\,ns).

\section{Dark counts measurements}

\subsection{Description of the method}

An original method has been design to measure the dark counts of the PMT without
prior knowledge of the photoelectron position peak, i.e, without prior
calibration. The ultimate goal of this method would be to provide an additional
calibration of the OD PMTs in Hyper-Kamiokande using all the data collected from
the ID triggers for example.

The method can be divided into two steps\,: first, an algorithm scan the
waveforms looking for dark pulses signals. Then, a histogram of the integrated
waveforms is constructed and fitted with a well known response function from
\cite{Bellamy:1994bv}. The innovation comes from the search of the dark pulses
signals.

We recreated a situation analog to the data taking at Hyper-Kamiokande\,: the ID
is replaced by our LED driver triggering at a frequency rate $f_\text{trigger}$,
in our case at the level of the MHz to increase statistics. The optical fibre is
not connected so there is no light emitted towards the PMT, but everytime a
trigger is emitted the signal waveform is recorded on the VME.

\begin{figure}[!htb]
  \begin{minipage}[c]{0.49\textwidth}
    \centering
    \includegraphics[width=1.\textwidth]{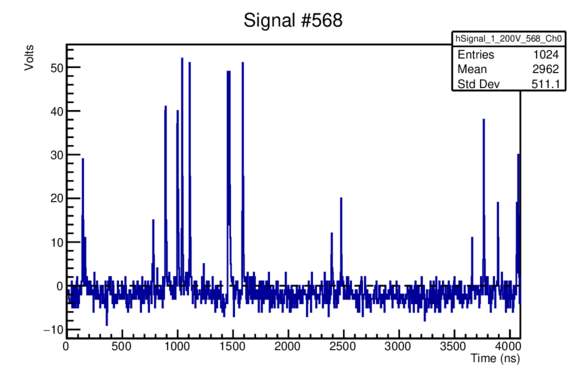}
  \end{minipage}\hfill
  \begin{minipage}[c]{0.49\textwidth}
    \centering
    \includegraphics[width=1.\textwidth]{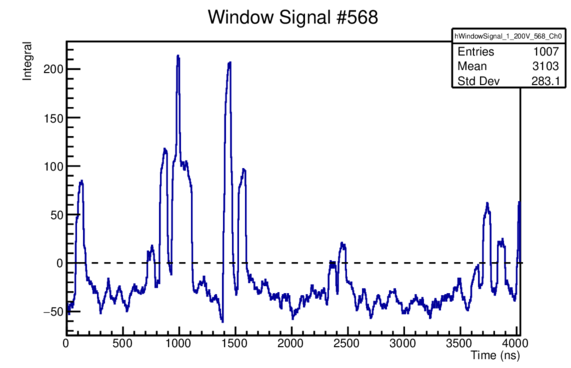}
  \end{minipage}
\caption{
  Left\;: Typical dark signal recorded from the Hamamatsu 3" PMT R14374.
  Right\;: Sliding window integration of same signal.
}
\label{signals}
\end{figure}

This data is sliced into bunches of $N_\text{bunch} = 64~\text{samples}$, and we
perform a sliding window integration on $N_\text{window} = 16~\text{samples}$
for each bunch. Then we fill an histogram where each bins is mapped with the
original starting position of the sliding window integration, and we extract the
ground baseline value of this signal $\mu_\text{dark}$, i.e the mean value on
the y-axis. This process is shown on Fig.\ref{signals} for an example signal
recorded from the Hamamatsu 3" R14374.

Therefore $\mu_\text{dark}$ can be used to defined an effective threshold for
dark pulses $\theta(\mu_\text{dark})$. In this proceeding we are using an affine
function, $\theta = \mu_\text{dark} + N\times\sigma_{\mu_\text{dark}}$ where
$\sigma_{\mu_\text{dark}}$ is the standard deviation for the distribution of the
$\mu_\text{dark}$ and $N$ an arbitrary parameter.

This procedures gives the threshold for the dark pulse charge $\theta$, which
can be now constructued from the data. Each pulses recorded is divided into same
bunches of $N_\text{bunch}$, and the sliding window integration is performed and
compared to $\theta$. If the sliding window integration is below $\theta$ for
the whole bunch, then the bunch is considered a pedestal event and integrated
over $N_\text{charge} = 32~\text{samples}$. If the sliding window integration is
above $\theta$, the bunch is considered containing a dark pulse event, where we
perform a constant fraction discriminator (CFD) algorithm to extract the
starting time of integration of the signal over $N_\text{charge}$.

This allows to construct the \emph{charge} histogram, for every integrated
pedestal and signal events. By using the fit function described in
\cite{Bellamy:1994bv}, we can extract the pedestal and dark pulse position to
compute the PMT gain. From the dark pulse width, we can infer the photoelectron
resolution, and from the dark pulse height the dark counts by normalizing to the
time of the acquisition.

\subsection{Results for the Hamamatsu 3" R14374 and the 3.5" R14689}

\begin{figure}[!htb]
  \begin{minipage}[c]{0.49\textwidth}
    \centering
    \includegraphics[width=1.\textwidth]{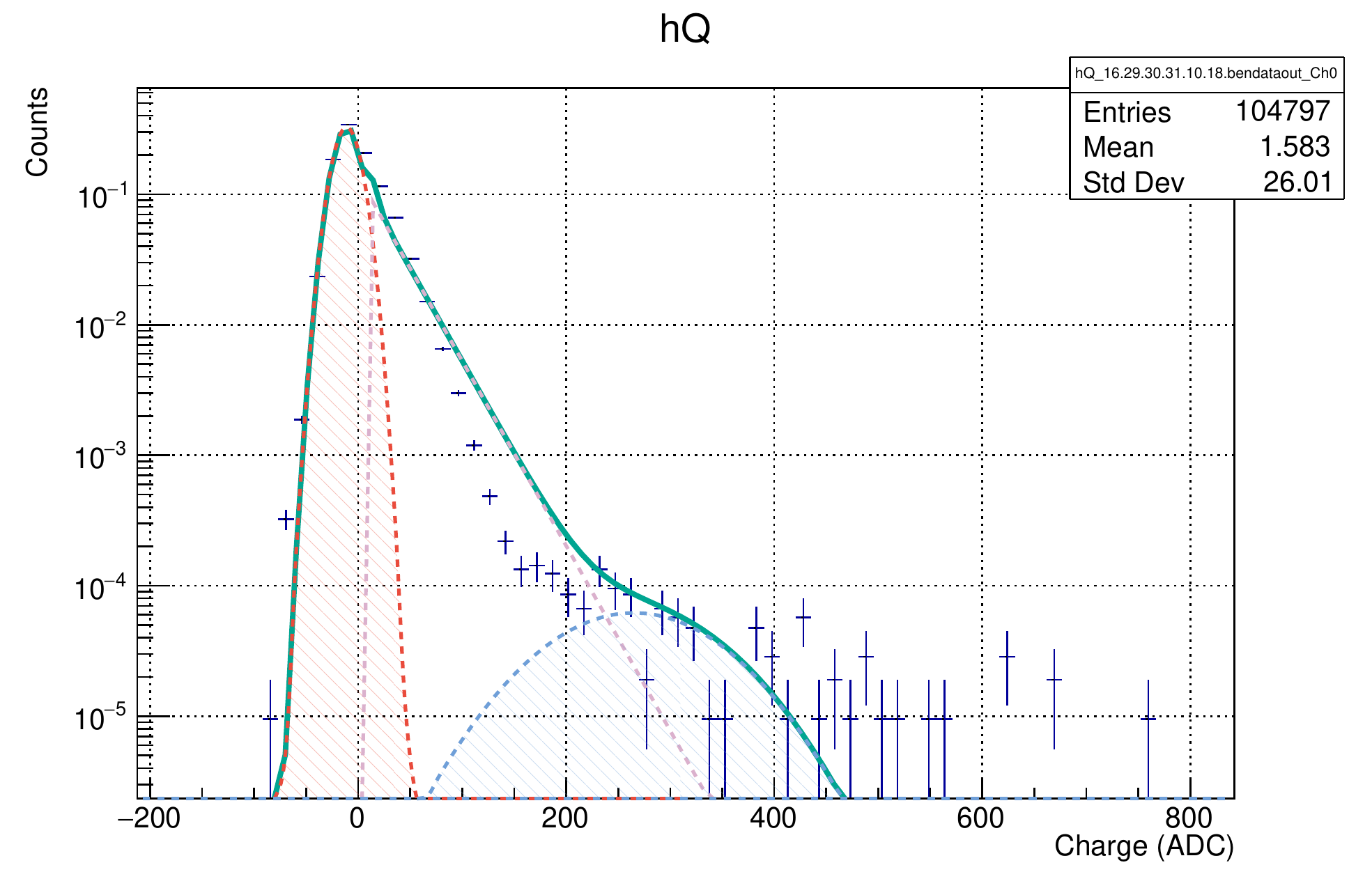}
  \end{minipage}\hfill
  \begin{minipage}[c]{0.49\textwidth}
    \centering
    \includegraphics[width=1.\textwidth]{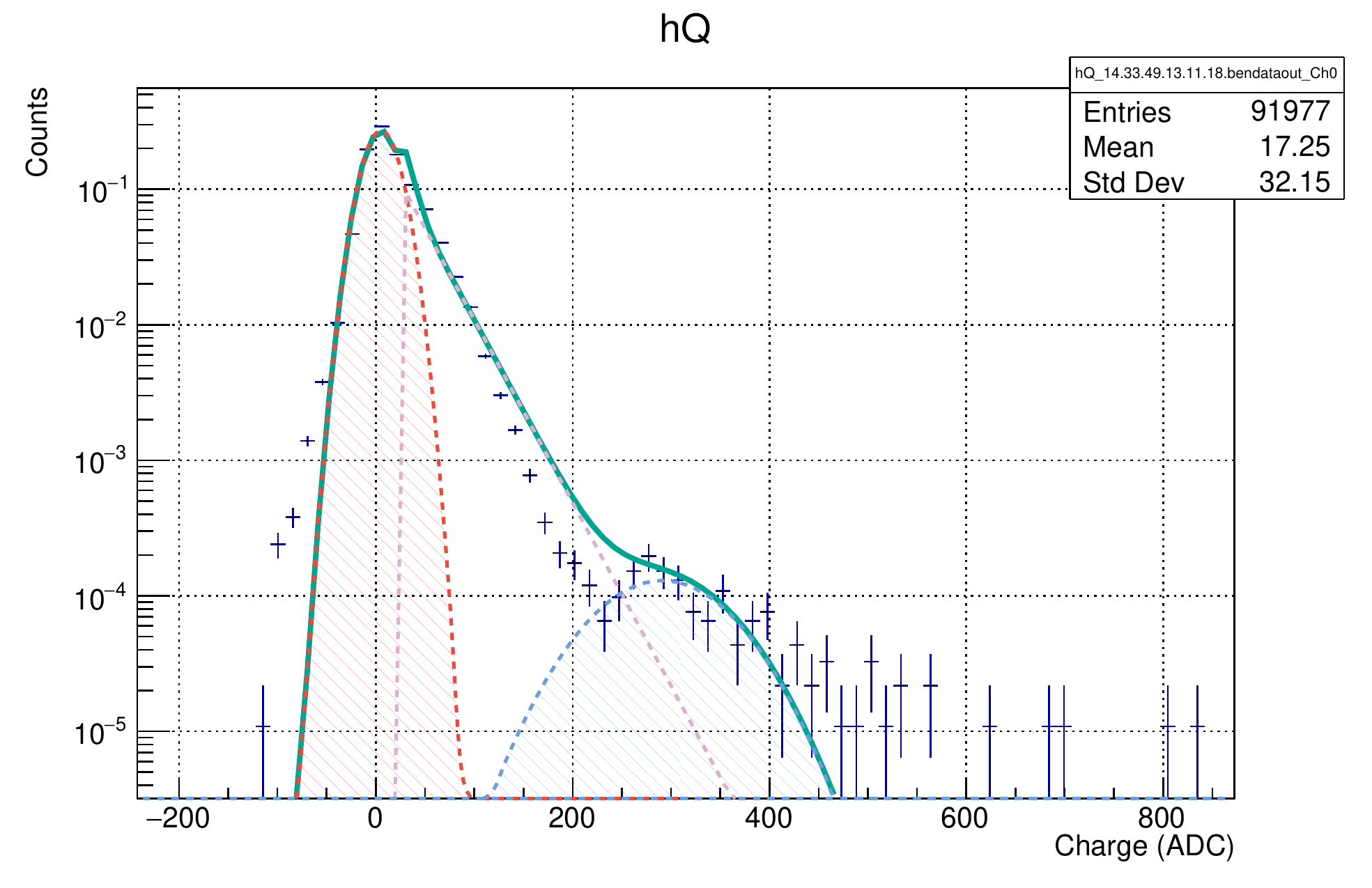}
  \end{minipage}
\caption{
  Left\;: Dark counts histogram from the Hamamatsu 3" PMT R14374.
  Right\;: Dark counts histogram from the Hamamatsu 3.5" PMT R14689.
  Both\,: The red and blue area correspond to the pedestal and dark counts hits respectively.
  In green, the fit is represented over the data.
}
\label{darkcounts}
\end{figure}

The Fig.\ref{darkcounts} shows the histogram constructed with the method
described in previous section, for the Hamamatsu PMTs R14374 (3") and R14689
(3.5"). The extracted parameters are summarized in Table.\ref{t1}.

\begin{figure}[!htb]
    \centering
    \includegraphics[width=0.49\textwidth]{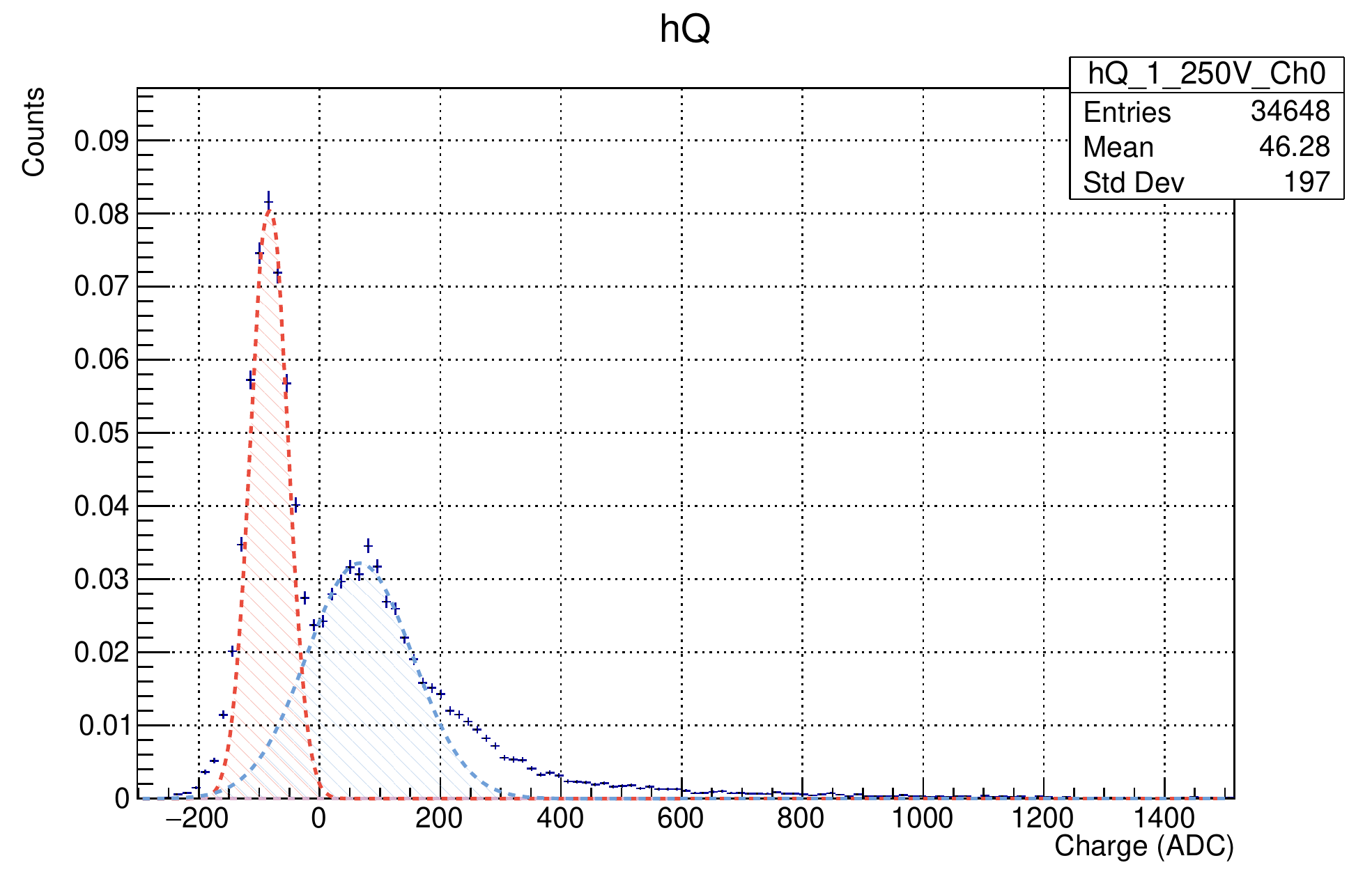}
\caption{
  Dark counts histogram from the Hamamatsu 3" PMT R14374, with prior calibration beforehand.
  The red and blue area correspond to the pedestal and the first photoelectron hits respectively.
}
\label{darkcalib}
\end{figure}

The consistency of the algorithm was checked by performing a similar analysis
but with calibration of the PMTs beforehand (Fig.\ref{darkcalib}). Knowing the photoelectron
threshold, the same charge histogram have been constructed and adjusted with the
same response function, yielding to according results. The errors quoted
correspond to the fit parameters errors, and are smaller when the PMTs are
calibrated beforehand. This effect is under investigation, but hints that a
possible measurement of the PMTs dark rates without calibration can be accurate.

\begin{table}[!htb]
 \caption{
 Summary of parameters characterized for Hamamatsu R14374 and R14689 PMTs.
 }
 \label{t1}
 \begin{tabular}{llll}
  \hline
  \hline
  \textbf{Model} & \textbf{Gain} & \textbf{Dark rates} & $\sigma_\text{SPE} / \mu_\text{SPE}$\\
  \hline
  \hline
  R14374 (3")   & $2.7\pm0.1\times10^6$ & $210\pm80$ Hz & $30\pm10\%$ \\
  \hline
  R14689 (3.5") & $2.8\pm0.1\times10^6$ & $250\pm100$ Hz & $17\pm11\%$ \\
  \hline
 \end{tabular}
\end{table}

\section{Conclusion}

The Hyper-Kamiokande outer-detector will consist of an array of 13.3\,k 3" PMTs
facing outwards and signaling the presence of charged particle by clusters of
hits above a certain threshold. The choice of small tubes for Hyper-Kamiokande
is motivated by the need for a finer array of detectors to veto the small
thickness of the barrel region, and by the lower cost and dark rate of the
smaller tubes.

A new method to measure the dark counts of small 3" PMTs without prior
calibration has been developped and tested on two Hamamatsu PMTs, which are
candidates to be installed in the Hyper-Kamiokande outer-detector. The measured
dark counts listed in Table.\ref{t1} is in agreement with the specifications for
the outer-detector and is being compared to two other models from Electron Tubes
(ET9320KFL) and from HZC (XP82B20).


\begin{thebibliography}{9}
  \bibitem{Abe:2018uyc}
  K.~Abe \textit{et al.} [Hyper-Kamiokande Collaboration],
  arXiv:1805.04163 [physics.ins-det].
  \bibitem{Fukuda:2002uc}
  Y.~Fukuda \textit{et al.} [Super-Kamiokande Collaboration],
  Nucl.\ Instrum.\ Meth.\ A \textbf{ 501}, 418 (2003).
  \bibitem{Aiello:2018nvl}
  S.~Aiello \textit{et al.} [KM3NeT Collaboration],
  JINST \textbf{13}, no. 05, P05035 (2018).
  \bibitem{Bellamy:1994bv}
  E.~H.~Bellamy \textit{et al.},
  Nucl.\ Instrum.\ Meth.\ A \textbf{339}, 468 (1994).
\end{thebibliography}
\end{document}